\documentclass[%
twocolumn,
superscriptaddress,
nofootinbib,
 amsmath,amssymb,
 aps, prd
]{revtex4-2}

\usepackage{graphicx}
\usepackage{tensor}
\usepackage{siunitx}
\usepackage{xcolor} % Remove after notes are no longer needed
\usepackage[normalem]{ulem} % Remove after notes are no longer needed
\usepackage{lipsum} % Remove after filler text is no longer needed

\usepackage[pdfborder={0 0 0}, plainpages=false, hyperfigures=true]{hyperref}

\newcommand{\beq}{\begin{equation}}
\newcommand{\eeq}{\end{equation}}
\newcommand{\beqn}{\begin{eqnarray}}
\newcommand{\eeqn}{\end{eqnarray}}

\newcommand{\lo}{\mathrel{\raise.3ex\hbox{$<$}\mkern-14mu
    \lower0.6ex\hbox{$\sim$}}}
\newcommand{\go}{\mathrel{\raise.3ex\hbox{$>$}\mkern-14mu
    \lower0.6ex\hbox{$\sim$}}}

\newcommand{\UNH}{\affiliation{Department of Physics \& Astronomy, University of New Hampshire, 9 Library Way, Durham NH 03824, USA}}

\begin{document}
\title{Disk mass predictions for binary neutron star mergers: limitations of proposed symbolic regression models}
\author{Francois Foucart}\UNH

\begin{abstract}
Modeling disk formation and mass ejection in binary neutron star systems is an important component in the construction of models for the electromagnetic signals powered by these events. Most models rely on analytical formulae for the disk mass and dynamical ejecta that are fitted to the results of numerical simulations, yet these fits have large uncertainties that significantly limit our ability to extract information from merger observations. In a recent manuscript, Darc {\it et al} claim that disk mass formulae constructed using symbolic regression outperform existing formulae and robustly extend to regions of the parameter space outside of the fitting region. I show here that the improvement over the most directly comparable existing model comes mostly from the use of different error measures in optimizing the fitting parameters. For the limited training data used so far, that existing fitting formula has a performance similar to symbolic regression models when optimized over the same error measure. More importantly, I show that many of the formulae obtained through symbolic regression provide unphysical results when used over the whole range of parameters relevant to the modeling of binary neutron star mergers, making them dangerous to use within parameter estimation pipelines. I conclude that fitting formulae with more physics input (e.g. Lund {\it et al} 2025), albeit certainly imperfect, remain safer to use in data analysis than these symbolic regression results. Symbolic regression results used in conjunction with careful physics-based vetting may however outperform them in the future.
\end{abstract}

\maketitle

\section{Introduction}

Electromagnetic signals form binary neutron star mergers carry in theory a wealth of information about the properties of the merging compact objects, and the equation of state of dense nuclear matter. Extracting that information remains however a difficult task. An important step in that process is to model the mass, composition and geometry of the matter ejected by mergers as a function of binary parameters. This is typically done through fitting formulae calibrated on the results of numerical simulations. In particular, fitting formulae exist for the mass ejected during merger~\cite{Radice:2018pdn,Kruger:2020gig,Nedora:2020qtd}, the mass of the accretion disk produced by a given merger~\cite{Dietrich:2016fpt,Radice:2018pdn,Kruger:2020gig,Nedora:2020qtd,2025ApJ...987...56L}, the fraction of that disk that is eventually unbound~\cite{Raaijmakers:2021slr}, and the composition and scale height of the ejecta~\cite{Nedora:2020qtd}. These formulae are known to come with very large uncertainties, especially far from the best simulated regions of parameter space~\cite{Henkel:2022naw}. Part of that uncertainty is due to the inhomogeneous nature of available numerical data~\cite{Nedora:2020qtd}, with simulations differing in both numerical accuracy and level of microphysics modeling. A second issue is the uneven coverage of the available parameter space. As a result, fitting formulae using different functional forms and/or different reference numerical simulations show very large differences in their predictions~\cite{Henkel:2022naw}. Finally, many fitting formulae have only partial physical motivation. Their functional forms are derived through a combination of physical intuition and trial and error; or as simple polynomial expansions.

Recently, Darc {\it et al}~\cite{Darc:2025set} proposed symbolic regression as an alternative. In symbolic regression, the functional form of the fitting formulae is itself optimized over, given a set of allowed mathematical operations. This is in theory an interesting option, removing some of the arbitrariness and guesswork used in current methods. Darc {\it et al}~\cite{Darc:2025set} presents symbolic regression models which are claimed to exceed the performance of existing models, to allow models to capture the dependance of the disk mass in more parameters, and to robustly extend to regions of the parameter space outside of the fitting region. While recognizing the potential of symbolic regression for future work, I respectfully disagree with these conclusions.

First, I will show here that most of the improvement over the most directly comparable existing model (i.e. our model from~\cite{Kruger:2020gig} calibrated on the same training set) comes from measuring the quality of the fit using a different metric from the one used to optimize the original model. The symbolic regression models are then `better' according to the metric they were optimized for, but comparable performance can be obtained from the old model if it is optimized using the same metric for success. Additionally, and more importantly, we will see that the `best' model found by symbolic regression is less reliable outside of the range of parameters where it is trained than models with mode physics inputs, including~\cite{Kruger:2020gig} and the more recent model from~\cite{2025ApJ...987...56L}. Most of the symbolic regression models show unphysical results for binaries with neutron stars more compact than the calibration set -- a range of parameters that is definitely important for parameter modeling. This shows the importance of carefully assessing the limitations of models with largely arbitrary functional forms, and how these functions behave within the parameter space that we wish to model. Finally, I note that the potential of higher-dimensional models to be useful for parameter estimation is debatable. There is no doubt that the use of additional parameters will eventually be needed. Yet the `best' model of Darc {\it et al}~\cite{Darc:2025set} is still a single-parameter model, and in cases where multi-parameters models overperform single-parameter models in terms of their mean square error, that improvement disappears when considering metrics that account for model complexity (e.g. the Bayesian Information Criteria). Overall, I conclude that while existing models certainly have significant uncertainties in their prediction, the more physics-based model of Lund {\it et al}~\cite{2025ApJ...987...56L} that was calibrated to the most recent numerical data likely is much safer to use in parameter estimation pipelines than the symbolic regression models.

In the following section, I briefly discuss the numerical data used in all fits. I then discuss the performance of the best single-parameter models, and show that the better performance of symbolic regression models shown in~\cite{Darc:2025set} is mainly due to the metric chosen to quantify errors. I then consider the behavior of these single-parameter models over the physically acceptable parameter space for binary neutron stars, and show that most symbolic regression models perform poorly out of their fitting region -- and that the one that does not perform poorly is in practice a close match to the results that would be obtained by refitting an existing model on the same error metric.

\section{Numerical data and Simulated Parameter Space}

The main objective here is to produce models for the mass of the accretion disk formed after a binary neutron star merger, $M_{\rm disk}$~\cite{Kruger:2020gig,Nedora:2020qtd}. We first note that the very definition of $M_{\rm disk}$ is ambigous. For systems that form a neutron star surrounded for a disk, a common choice is the mass of bound matter at density $\rho < 10^{13}\,{\rm g/cm^3}$. For black hole-disk remnants, $M_{\rm disk}$ is tyipcally defined as the bound matter outside the black hole. Both quantities vary over time, and different simulations unfortunately report values at different times. When fitting to numerical data, we would ideally account for this, and other uncertainties in the simulation results. These include pure numerical error, i.e. the error in the simulations assuming that they perfectly model a binary neutron star system; as well as the impact of missing physics in simulations. The former has only been quantified for a subset of simulations ($\sim 10\%$ relative errors are common). The second is potentially more significant, and not easily quantified. In~\cite{Kruger:2020gig}, we used
\beq
\Delta M_{\rm disk} = 0.5 M_{\rm disk} + 0.0005M_\odot.
\label{eq:err}
\eeq
The $0.0005M_\odot$ absolute error was introduced because lower disk masses are likely not resolved in existing simulations. The $50\%$ relative error was, admittedly, a largely arbitrary choice based solely on the fact that larger disk mass tend, in our experience, to lead to larger absolute errors. Multiplying $\Delta M_{\rm disk}$ by a constant does not change best fit values in the models, so only the ratio of the relative and absolute errors has an impact on fitting results. 

As in~\cite{Darc:2025set}, I consider here models fitted to the 56 simuations used in~\cite{Kruger:2020gig} (training set), and then tested on the broader range of simulations used by~\cite{Nedora:2020qtd} (testing set; with 119 simulations, including the 56 from~\cite{Kruger:2020gig}). The training dataset has binary mass ratios in the range $q\in  [0.77-1]$, while the testing dataset has $q\in [0.55,1]$. The main parameter used in our fit is the compactness of a neutron star $C=GM/(Rc^2)$, with $(M,R)$ the gravitational mass and Schwarzschild radius of a neutron star. Specifically, based on the consideration that a large part of the ejecta comes from the tidal disruption of the lower mass star, we will rely on the compactness of that lower mass star $C_1$. The fact that $C_1$ appears to be the parameter that most directly correlates with the fitting data was the motivation behind the model of~\cite{Kruger:2020gig}. That assumption was confirmed by both the fits on a larger dataset performed in~\cite{2025ApJ...987...56L} and the symbolic regression results of~\cite{Darc:2025set}. The training data has $C_1\in [0.135,0.205]$. The testing data has $C_1\in [0.120, 0.205]$. We note that the upper bound on $C_1$ is a limitation of the numerical datasets. A neutron star of mass $2M_\odot$ and radius $10\,{\rm km}$ would have $C\sim 0.3$. A $1.7M_\odot$ neutron star of the same radius would have $C\sim 0.25$. The former is unlikely to be the `lower mass star' in a binary neutron star system, but the latter is fully consistent with the lower mass object in GW190425~\cite{LIGOScientific:2020aai}. One of the reasons no simulations exist at high $C_1$ is that systems with $C_1>0.2$ are likely to be of high total mass and lead to the rapid collapse of the merger remnant to a black hole, with no disk formation -- a less interesting system to model than lower mass binaries. Nonetheless, if a fitting formula is to be used for data analysis, it should avoid predicting large $M_{\rm disk}$ in that regime. The lower bound is more realistic: a $1.1M_\odot$ neutron star with $R=14\,{\rm km}$ has $C\sim 0.115$, a fairly reasonable lower limit.

\section{Models depending on the lower star's compactness}

In~\cite{Darc:2025set}, the quality of models is estimated using both the Mean Square Error (MSE) and the Bayesian Information Criteria. In both cases, the best model found by symbolic regression is a model depending only on $C_1$:
\beq
M_{\rm disk}^{\rm SR} = 0.118824-(0.142985  \sin{(\sin{(40.896317  C_1)})}).
\eeq
We will largely focus on such one-parameter models here. As discussed in the introduction, there is currently no indication that multi-parameter models perform better on existing data.

The MSE of the best symbolic regression model is $\sim 0.0027 M_\odot^2$. In~\cite{Darc:2025set}, this error is compared to a few existing fitting formulae; specifically the model of~\cite{2025ApJ...987...56L}, which depends only on $C_1$ and was fitted to a larger set of simulations; our model from~\cite{Kruger:2020gig}, which depends on $C_1,M_1$ (with $M_1$ the mass of the lower mass neutron star; a parameter which enters only as a global linear scaling for the model) and was fitted to the same training set as the symbolic regression model; and the model of~\cite{Radice:2018pdn}, which used a smaller training set and depends on the tidal deformability of the system. All these models have higher MSEs than the symbolic regression model ($0.0029M_\odot^2$ for~\cite{2025ApJ...987...56L}; $0.0037M_\odot^2$ for~\cite{Kruger:2020gig}; and $0.0062M_\odot^2$ for~\cite{Radice:2018pdn}). That the models performed better with more training data is expected. From these results, it might also seem that when using the same training set, the symbolic regression model performed better on the testing set.

There is however an important caveat to those results: different fitting methods use different error measures to optimize the fitting parameters. We illustrate this point using our own model from~\cite{Kruger:2020gig}. In that manuscript, we minimized the reduced $\chi^2$ of the fit assuming the errors from Eq.~\ref{eq:err}. This is very different from using the MSE, which effectively assumes constant absolute numerical errors. Specifically, a model assuming the errors from Eq.~\ref{eq:err} will weight more heavily simulations producing low-mass disks, and very heavily simulations producing nearly no disk.

To make comparisons with~\cite{Darc:2025set} and plotting of the results easier, let us consider a slight modification of~\cite{Kruger:2020gig}
\beq
M_{\rm disk}^{\rm K20} = \max\left(\left(aC_1+b\right)^d,0.0005\right) M_\odot.
\eeq
The original formula differs from this one by a factor of $M_1$ only. {\it If we optimize the fit for the MSE} and fit only the training set, we find $a=-4.47115082$, $b=0.7851083$, $d=0.92839854$ and an MSE of $0.0030M_\odot^2$ on the full testing set. In~\cite{Darc:2025set}, three different symbolic regressions algorithm were used to create models depending only on $C_1$.  This simple refitting nearly exactly matches the average performance of the symbolic regression results (not even accounting for the fact that symbolic regression optimizes over a larger volume of possible models). If on the other hand we optimize the fit to minimize $\chi^2$ using Eq.~\ref{eq:err} as our error estimate, we find a MSE of $0.0034M_\odot^2$ on the testing set. Does this make the latter a worse model? It depends on our definition of `worse'. Its MSE is $20\%$ higher than the best symbolic regression model, and higher than the MSE of all three single-parameter symbolic regression models from~\cite{Darc:2025set}; but its $\chi^2$ on the full data set assuming the errors of Eq.~\ref{eq:err} is $7$ times smaller than the best full symbolic regression model! Unsurprisingly, each model is best when using the metric for which it was optimized... and there are no strong reasons to prefer either error measure.

\begin{figure}
\includegraphics[width=0.9\columnwidth]{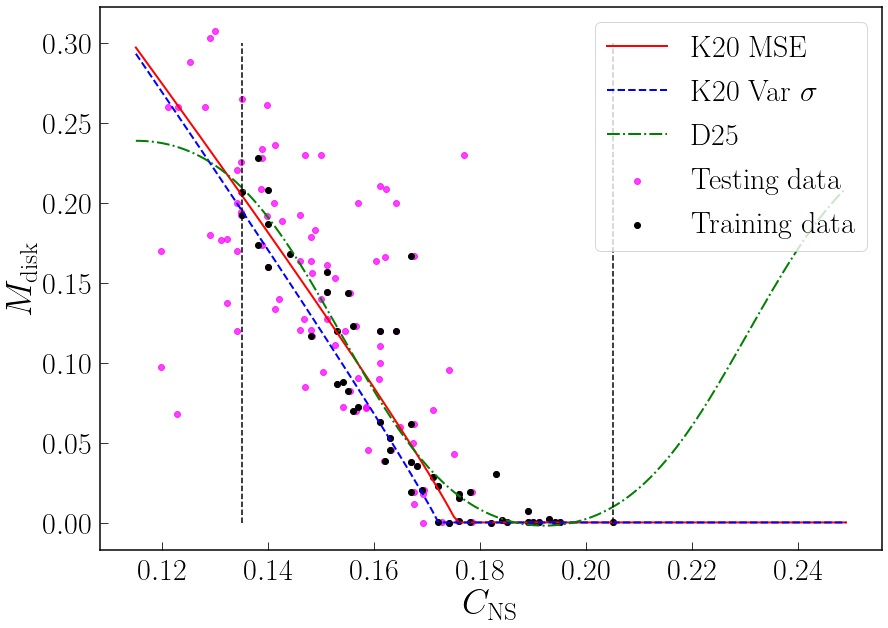}
 \caption{Behavior of three one-parameter models for $M_{\rm disk}$. We show the best symbolic regression model from~\cite{Darc:2025set} (D25), and the two models derived here by minimizing the MSE (K20 MSE) and by using the errors from Eq.~\ref{eq:err} (K20 Var $\sigma$), based on our previous model~\cite{Kruger:2020gig}. We use black circles for the training data, and gray circle for the testing data. Vertical dashed lines show the range of compactness in the training set.}
\label{fig:fit}
\end{figure}

Fig.~\ref{fig:fit} shows the best symbolic regression model from~\cite{Darc:2025set}, and the two models derived here by optimizing either the MSE or $\chi^2$, together with the training and testing datasets. We show results for $C_1 \in [0.115,0.25]$, fairly conservative bounds on the range of $C_1$ that should be modeled for the purpose of parameter estimation on actual observations. Within the range of the training dataset, all three models are unsurprisingly in good agreement. The model minimizing $\chi^2$ puts more emphasis on capturing low disk mass results, as expected. Fig.~\ref{fig:fit} shows that the MSE on the testing set is strongly impacted by two points with low $C_1$ and low $M_{\rm disk}$ in the bottom-left corner of the plot, for all models. Those are the two highest mass ratio systems in our dataset ($q=2$ and $q=5/3$)\footnote{There is a second system with $q=5/3$, with $C\sim 0.13$ and $M_{\rm disk}\sim 0.14$; that system is also poorly modeled by all fitting formulae, though not as poorly as the other two}. The weight given to those systems in the error budget is largely arbitrary -- if we had $20$ such systems in the testing data instead of $2$, our MSE would be significantly larger; if we had none, it would be smaller. The correct weight to give those systems would ideally be set by population models, yet existing population models are certainly not good enough for us to do this explicitly. All we can say is that all of the models become worse once the mass ratio increases beyond the range simulated in the training set. A model that depends on $q$ is certainly needed to capture those data points.

It should thus be clear from Fig.~\ref{fig:fit} that the main source of error in these models is the dependence of $M_{\rm disk}$ in parameters beyond $C_1$, rather than the quality of the single-parameter fit. However, no model using other parameters appears to perform better for the limited dataset considered here, indicating that this training set is likely insufficient to capture the impact of these other parameters. It is also worth noting that, considering the sparse and uneven coverage of the parameter space in the testing set, $10\%-20\%$ variations in the MSE on the testing sets are not very meaningful: a single additional simulation comparable to the two main outliers would be enough to change the MSE by $\sim 10\%$.

\section{Model robustness}

\begin{figure}
\includegraphics[width=0.9\columnwidth]{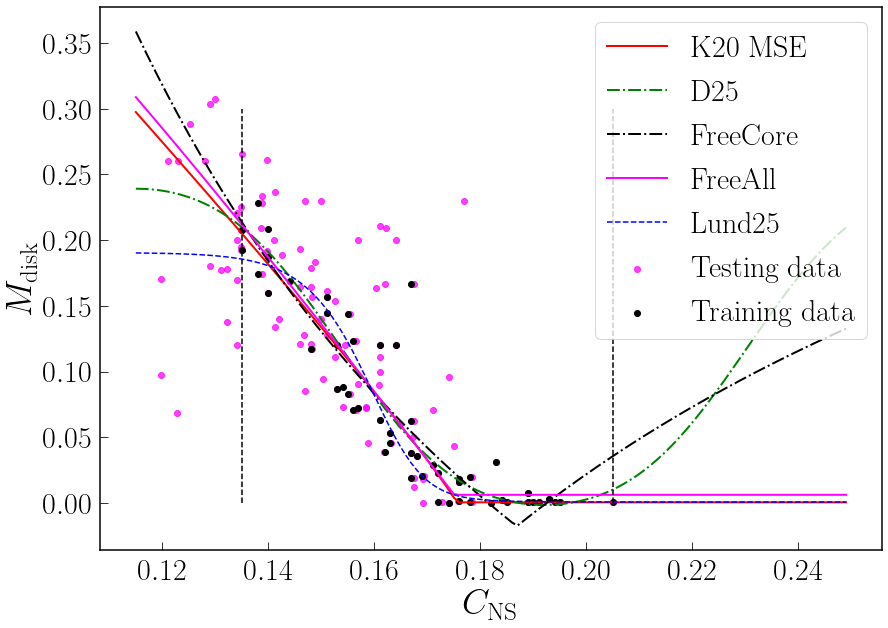}
 \caption{Same as Fig.1, but now for all models from~\cite{Darc:2025set} (D25) that only depend on $C_1$, as well as the model derived here by minimizing the MSE (K20 MSE) and the existing model from~\cite{2025ApJ...987...56L}.}
\label{fig:fit2}
\end{figure}

In the previous section, we focused rather narrowly on errors measured on the testing dataset.  This is not sufficient to address the robustness of the models, because the testing set itself is severely limited. Fig.~\ref{fig:fit} shows how the models considered in the previous section extrapolate to higher and lower compactness. Fig.~\ref{fig:fit2} shows similar results for all one-parameter models of~\cite{Darc:2025set} (D25 -- called PyOp in~\cite{Darc:2025set}, FreeCore, FreeAll), as well as the recent fitting formula derived without symbolic regression by Lund {\it et al}~\cite{2025ApJ...987...56L}. The easiest issue to address is the high-compactness results. There, we expect $M_{\rm disk}$ to nearly vanish. The models from~\cite{Kruger:2020gig,2025ApJ...987...56L} are explicitly constructed to do this, and unsurprisingly recover the correct result. The best symbolic regression model behaves in a clearly unphysical manner. This is not surprising for a model without physical input. The behavior of the symbolic regression model at high compactness clearly belies the claim that the model is `more robust' than existing formulae outside of its training range.

At low compactness, and within the range of $C_1$ considered here, results are more ambiguous. The K20 models fits better results for near equal mass binaries. The symbolic regression model fits (slightly) better the two outlier points at high mass ratio. The Lund model~\cite{2025ApJ...987...56L} clearly leans even more towards the asymmetric mass ratio results, and performs worse on near equal mass systems. We note however that, for the symbolic regression model, good agreement will only be observed over a very narrow range of compactness. The D25 model reaches a maximum around $C_1=0.115$ and then oscillates back to zero -- an unphysical result. It is clear that the use of a sine function is causing major issues in terms of the physical realism of the solution. The `best' symbolic regression model would be very dangerous to use within any parameter estimation pipeline, much more so that the models from~\cite{Kruger:2020gig,2025ApJ...987...56L}. At very low compactness, the model from~\cite{Kruger:2020gig} would eventually predict unphysically large values of $M_{\rm disk}$ -- though only well outside of the range of compactness relevant to neutron star mergers. The model from~\cite{2025ApJ...987...56L} simply asymptotes to $0.2M_\odot$.

We note that this issue is not encountered by all symbolic regression models. In Fig.~\ref{fig:fit2}, we see that the `FreeAll' model of ~\cite{Darc:2025set} extrapolates outside of the training range in the same way as our model. That model is however nearly identical to our model -- with a more complex functional form and a slightly higher MSE. It is interesting that the one symbolic regression model that does extrapolate somewhat reasonably out of its training range of compaction effectively reproduces the result of an existing fitting formula. While that does not represent an improvement over the state-of-the art, it does point to the fact that with careful choices of allowed functional forms and vetting of the physical realism of symbolic regression models, they could indeed facilitate work done to model $M_{\rm disk}$, especially once more complete datasets allows us to develop reliable multi-parameter models.

\section{AI-driven symbolic regression}

We note that in an attempt to get a more `physics inspired' model, Darc {\it et al}~\cite{Darc:2025set} also asked ChatGTP to generate the basic format of a symbolic regression model that would capture the main known physical behavior of neutron star binaries, including the fact that prompt collapse to a black hole. This was done in part by feeding our paper discussing existing disagreement between models~\cite{Henkel:2022naw} to ChatGPT. While ChatGPT did identify some of the important trends that one would expect in $M_{\rm disk}$ on physical ground (mixed with some decidedly irrelevant comments), the suggested form of the expression to optimize with symbolic regression appears to be a Frankenstein monster of all binary parameters that might be relevant, without any actual physical input. The final model after symbolic regression as reported in~\cite{Darc:2025set} only depends on two parameters however:
\beqn
M_{\rm disk} &=& 18170.047 C_1^{6.911347}- 6.031138 \exp{(C_1)} \nonumber\\
&& - 0.02342195 M_2 - 0.017214041 \times 432.73465 \nonumber\\
&& - 0.3142357.
\eeqn
This returns large negative $M_{\rm disk}$ values for all physical values of $(C_1,M_2)$ that I tried. Which is fitting for ChatGPT, I suppose. Even if there is a simple printing error in the functional form in the model, this expression looks decidedly far from anything `physics inspired'.

\section{Conclusions}

While physics-informed uses of symbolic regression to fit the results of numerical simulations is an interesting avenue to pursue if enough reliable data is available, the dataset used in~\cite{Darc:2025set} does not appear to offer enough information to do more than confirm the dominant dependence of the disk mass results on the compactness of the lower mass star $C_1$. The best single-parameter symbolic regression models have performance comparable to standard numerical fits within the fitting window, and their sometimes peculiar choice of functional forms for the fitting formulae (sines of sines, $e^{C}$) lead to unphysical behavior outside of their fitting range; more so at least than for existing models with a moderate amount of physical motivation.

Symbolic regression models could nonetheless be more reliable outside of their fitting region with more physics-based inputs and/or with careful vetting for physically reasonable behavior. It is notable for example that one of the symbolic regression models from~\cite{Darc:2025set} -- the only one extrapolating well to high compactness -- is nearly identical to an existing model when calibrated on the same data and for the same error measure. In one dimension, this may not be overly helpful -- it is fairly easy to come up with a reasonable functional form for a 1D model by inspecting the data, especially when we can only hope to get order-of-magnitude agreement between the model and the data. Applying symbolic regression to an expanded training set with more high mass ratio systems, on the other hand, may lead to faster selection of an appropriate functional form for higher dimensional models -- and in that respect, the results of~\cite{Darc:2025set} give hope that symbolic regression could help reduce the time needed to construct models through trial and error in the future.

Until such models are built, however, the results presented here indicate that the safest bet for using model within parameter estimation pipeline remains the use of existing fitting formulae. In particular, the model from Lund {\it et al}~\cite{2025ApJ...987...56L} is calibrated on very recent numerical data (a more extended training set than what we use here), behaves correctly at high $C_1$ (where the correct physical result is known), and at least reasonably at low $C_1$ (where the spread in the possible values of $M_{\rm disk}$ for a given $C_1$ is too large for one-parameter models to be reliable anyways). Considering the large disagreements between existing models in extreme regions of the binary neutron star parameter space, an even better process is likely to perform any parameter estimation with multiple models, and consider the spread of resulting predictions as an estimate of modeling uncertainty~\cite{Henkel:2022naw}.

\begin{acknowledgments}
F.F. gratefully acknowledges support from the Department of Energy, Office of Science, Office of Nuclear Physics, under contract number DE-SC0020435; from the NSF through award AST-2510568; and from NASA through grant 80NSSC22K0719. 
\end{acknowledgments}

\bibliography{bibliography.bib}

\end{document}